\documentclass[sigconf]{acmart}
\usepackage{pifont}
\usepackage{pgfplots}
\usepackage{tikz}
\usepackage{caption}
\pgfplotsset{compat=1.18}
\usepackage{algorithm}
\usepackage{algpseudocode}
\usepackage{multirow}
\usetikzlibrary{calc}
\usepackage{adjustbox}
\usepackage{algpseudocode}
\usepackage{array}
\usepackage{balance}
\usetikzlibrary{patterns}
\usepackage{amsmath,amsfonts}
\usepackage{xcolor}
\usepackage{graphicx}
\usepackage{textcomp}
\usepackage{xcolor}
\usepackage{caption}
\AtBeginDocument{%
  }

\setcopyright{acmlicensed}
\copyrightyear{2026}
\acmYear{2026}
\acmDOI{XXXXXXX.XXXXXXX}
\acmConference[DAC '26]{Proceedings of the 63rd Design Automation Conference}{July 26--29, 2026}{Long Beach, CA, USA}
\acmISBN{978-1-4503-XXXX-X/2026/07}

\begin{document}

\title{PDF: PUF-based DNN Fingerprinting for Knowledge Distillation Traceability}

\author{Ning Lyu, Yuntao Liu, Yonghong Bai, and Zhiyuan Yan}
\begin{abstract}
  Knowledge distillation transfers large teacher models to compact student models, enabling deployment on resource-limited platforms while suffering minimal performance degradation. However, this paradigm could lead to various security risks, especially model theft. Existing defenses against model theft, such as watermarking and secure enclaves,  focus primarily on identity authentication and incur significant resource costs. Aiming to provide \emph{post-theft accountability and traceability}, we propose a novel fingerprinting framework that superimposes device-specific Physical Unclonable Function (PUF) signatures onto teacher logits during distillation. Compared with watermarking or secure enclaves, our approach is lightweight, requires no architectural changes, and enables traceability of any leaked or cloned model. Since the signatures are based on PUFs, this framework is robust against reverse engineering and tampering attacks. In this framework, the signature recovery process consists of two stages: first a neural network-based decoder and then a Hamming distance decoder. Furthermore, we also propose a bit compression scheme to support a large number of devices. Experiment results demonstrate that our framework achieves high key recovery rate and negligible accuracy loss while allowing a tunable trade-off between these two key metrics. These results show that the proposed framework is a practical and robust solution for protecting distilled models. 
\end{abstract}




\maketitle

\section{Introduction}
Knowledge distillation \cite{distillingknowle}  has become a key technique in machine learning, allowing a large, high-performing teacher model to transfer its knowledge to a smaller student model. By training with the teacher’s soft outputs, the student maintains strong generalization performance at a fraction of the computational cost. This makes distillation particularly valuable when deploying models on the hardware. The teacher model plays an important role as the true source of knowledge, data curation, architecture design, and training. In practice, especially on edge devices like FPGAs, IoT systems, or embedded platforms, teachers are often deployed locally to enable fine-tuning, continual learning, and real-time distillation—supporting adaptive AI applications without persistent cloud dependency.

However, implementing teacher models on hardware raises significant security risks. Once deployed, they are vulnerable to model extraction \cite{learningmodels}, reverse engineering\cite{reverse}, side-channel probing, or unauthorized replication. Critically, once a teacher model is stolen, attackers can distill new student models or repackage and resell the teacher model without accessing the original training data.

Researchers have explored embedding identity information into models through fingerprinting and watermarking. For example, Uchida et al.~\cite{Uchida_2017} encoded digital watermarks into the weights of convolutional layers, while Adi et al.~\cite{USENIX} proposed backdoor-style signatures for ownership verification. However, these approaches are software-only and rely on access to the model's internal parameters or special triggers, and can be erased by fine-tuning or knowledge distillation. In summary, watermarking can prove ownership but cannot trace a leaked model to a specific device.

From the hardware perspective, CNN weights are obfuscated with Physical Unclonable Function (PUF) responses \cite{li2020puf} to lock models to specific hardware, while Xu \textit{et al}.~\cite{ISCAS} modulated normalization layers using PUF-based IDs for black-box ownership verification. These works show how PUFs have been applied for execution control and authentication in machine learning models. However, they do not address model leakage or theft, and none of them explore using PUFs to enable traceability of distilled models back to the leaking device.

We propose a lightweight fingerprinting framework that superimposes PUF-derived identity onto teacher model logits through structured perturbations during distillation. These perturbations are inherited by the student models, creating recoverable behavioral fingerprints that persist through training and deployment. Our method enables \emph{teacher-level traceability}: if an adversary steals a teacher model and uses it to distill new student models, the embedded fingerprint allows us to identify which teacher was compromised. Compared with watermarking~\cite{USENIX,zhang2018protecting}, our approach requires no special triggers or internal access.  Compared with enclaves or encryption~\cite{privacypreserving}, it avoids heavy runtime overhead. 

\noindent \textbf{The main features of the proposed framework are as follows:}  
\begin{itemize}
    \item \textbf{PUF-based logit fingerprinting.} We propose a novel framework that superimposes PUF-derived perturbations directly onto the teacher model’s logits. These perturbations are subtle, device-specific, and reliably inherited by the student models during knowledge distillation. To better reflect real hardware behavior, we incorporate controlled PUF bit-flip noise into the PUF keys during training and evaluation.

    \item \textbf{Two-stage signature recovery process.} We design a two-stage recovery process for PUF key recovery. The first stage uses a neural network trained on synthetic datasets, so direct access to the actual teacher or student models is not required. The second stage applies Hamming distance decoder for error correction. This ensures robust recovery under noisy and constrained conditions.  

    \item \textbf{Bit compression scheme.} To support a large number of devices, we introduce a bit compression scheme that compresses multiple PUF key bits into compact logit vectors, significantly improving fingerprinting capacity without additional overhead. 

\end{itemize}
 The experiments show that our framework achieves high recovery rates with negligible student model accuracy loss. Furthermore, our framework provides a tunable trade-off between fingerprint recovery and student accuracy, which makes the framework practical for real-world deployment, offering a lightweight and effective solution against distillation-based model theft.

\section{Background}
\subsection{Knowledge Distillation}
Knowledge distillation, introduced by Hinton \textit{et al}.~\cite{distillingknowle}, enables a smaller {student model} to learn from a larger, more accurate {teacher model} by using the teacher’s soft logits rather than relying on only hard class labels. In the machine learning, logits are the raw outputs of a network before the softmax activation.

The knowledge distillation process has two stages: the teacher is first trained on labeled data and then fixed; next, the student is trained with a {distillation loss} to mimic the teacher’s predictions. These soft outputs capture inter-class relationships and uncertainty, helping the student develop more discriminative and robust representations.
The teacher’s logit $\mathbf{z}_t$ and student’s logit $\mathbf{z}_s$ are processed through a temperature-scaled softmax function to produce the softened probability distributions $\mathbf{p}_t$ (from the teacher) and $\mathbf{p}_s$ (from the student):
\begin{equation}
\mathbf{p}_t = \text{Softmax}\left(\frac{\mathbf{z}_t}{T}\right), \quad \mathbf{p}_s = \text{Softmax}\left(\frac{\mathbf{z}_s}{T}\right),
\end{equation}
where $T > 1$ is the temperature parameter that smooths the output distributions, amplifying the small probabilities assigned to non-target classes. 
The most common loss function used for distillation is the Kullback–Leibler (KL) divergence~\cite{distillingknowle} between the softened teacher and student outputs:
\begin{equation}
    \mathcal{L}_{\text{KD}} = T^2 \cdot \text{KL}\left( \mathbf{p}_t, \| \, \mathbf{p}_s \right),
\end{equation}
where the $T^2$ term corrects for the gradient scaling effect of temperature. In addition to the KL divergence, a simpler alternative is to match the raw logits directly using Mean Squared Error (MSE)~\cite{MSE}:
\begin{equation}
\label{eq:mse}
    \mathcal{L}_{\text{MSE}} = \frac{1}{d} \sum_{i=1}^{d} \left( z_{t,i} - z_{s,i} \right)^2,
\end{equation}
where $d$ is the number of the classes. The loss function eliminates the need for softmax and temperature tuning. Overall, knowledge distillation provides a principled framework for transferring a model’s predictions.

\subsection{PUFs as Hardware Fingerprints}
\label{sec:puf}

PUFs leverage inherent physical randomness to secure AI on edge devices. Architectures such as {SRAM startup patterns}~\cite{sram}, {ring oscillators}~\cite{ro}, and {arbiter delay chains}~\cite{arb} generate unique and unclonable responses from manufacturing variations. PUFs also generate keys only when needed, avoiding permanent storage and reducing leakage risks. This hardware-rooted identity enables embedding device-specific signatures into machine learning models for ownership verification, provenance tracking, and activation gating without external key management.

However, while PUFs are engineered for stability, they are not perfectly deterministic. In practice, slight hardware-level variations caused by factors such as temperature changes, voltage fluctuations, and aging can occasionally can introduce 1\%–5\%  intra-device bit flips ~\cite{PUF_NOISE, variation, variation1}. 

 Another practical limitation is the \emph{effective key space}. although an $n$-bit PUF ideally supports $2^n$ responses, bit bias, correlation, and unstable positions reduce its effective entropy~\cite{instable}. For example, a 10-bit PUF may behave closer to an 8-bit identifier in practice, the actual number of distinguishable devices is closer to $2^8 =256$.

Both the noisy behavior and the reduced effective key space of PUFs are incorporated into training
and inference in our framework, which improves robustness and ensures that PUF-based fingerprints can be applied reliably in practical edge deployment scenarios.

\begin{figure*}[t]
\centering
\vspace{-5pt}
\includegraphics[width=0.85\textwidth, height=0.4\textwidth]{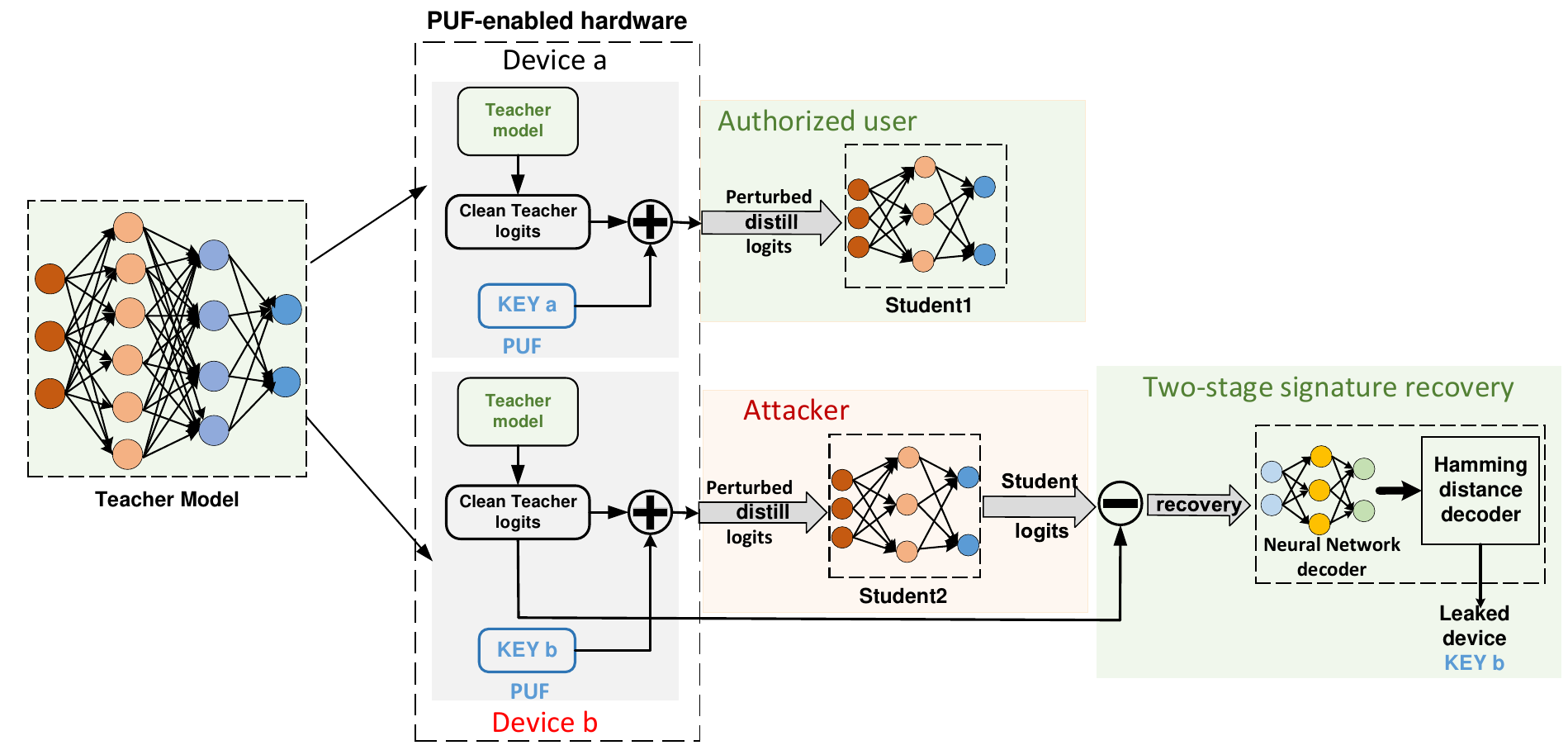}
\caption{The proposed PUF-based fingerprinting framework to defend against distillation-based model theft.}
\label{fig:system}
\end{figure*}

\section{Proposed Framework}
Our proposed framework consists of three major components: a PUF-based hardware fingerprinting scheme for the teacher model, a bit-compression technique for scalable key embedding, and a two-stage signature recovery process to identify the leaker device. Figure~\ref{fig:system} illustrates the fingerprinting and recovery flow and how they interact with the attacker.

\subsection{PUF-based Logit Fingerprinting}
In this section, we propose a method that superimposes PUF signals directly onto the logits of a trained teacher model. The key novelty of our approach is that these PUF-derived signatures persist in the behavioral outputs of student models, enabling any leaked or distilled model to be traced back to the originating device. This provides post-theft accountability.

Each hardware device is uniquely identified by a PUF key $\mathbf{k} \in \{0,1\}^n$, where $n$ denotes the bit length of the key (e.g., $n = 10$ or $50$). The effective PUF key space is smaller than $2^n$ due to bit bias, inter-bit correlation, and instability in practical implementations~\cite{instable} as discussed in Section~\ref{sec:puf}. To reflect this, the key space in our experiments is  a subset of $\{0,1\}^n$. We randomly sample a PUF key from the key space to represent the leaker's identity. 
To superimpose this identity onto the teacher model behavior, we introduce the perturbed logits $\mathbf{z}_{\text{PUF}}$ as
\begin{equation}
\mathbf{z}_{\text{PUF}} = \mathbf{z}_{\text{t}} + \boldsymbol{\delta},
\label{eq:perturbed_logits}
\end{equation}
where $\mathbf{z}_{\text{t}}$ denotes the clean teacher logits and $\boldsymbol{\delta}$ is a key-dependent perturbation vector. 
Specifically, the $i$-th component $\delta_i$ of $\boldsymbol{\delta}$ is obtained 
from the standard signed mapping, defined as
\begin{equation}
\delta_i = \epsilon \cdot (1 - 2 k_i),
\label{eq:delta_definition}
\end{equation}
with each bit $k_i \in \{0, 1\}$ mapped to a signed value in $\{-\epsilon, \epsilon\}$. 
Here, $\epsilon$ controls the strength of the fingerprinting.
This additive perturbation superimposes the PUF signal onto the teacher model output while preserving the model’s functionality.

The teacher network is quantized to $m$ bits (e.g., $m=8$) to represent realistic deployment constraints by using uniform quantization over inputs, weights and activations. Specifically, for a real value \( x \in [x_{\min}, x_{\max}] \), the quantized value \( \hat{x} \) is calculated as
\begin{equation}
\hat{x} = \frac{1}{S} \cdot \mathrm{round}\!\left(S \cdot \mathrm{clip}(x, x_{\min}, x_{\max})\right)
\label{eq:quantization}
\end{equation}
with scaling factor \( S = \frac{2^m - 1}{x_{\max} - x_{\min}} \). After quantization, the teacher model is trained (or fine-tuned) using categorical cross-entropy loss function. Once trained, the teacher model produces clean output logits $\mathbf{z}_{\text{t}}$.

A smaller student model is then trained by using perturbed logits $\mathbf{z}_{\text{PUF}}$ as soft supervision. Letting $\mathbf{z}_s$ represent the student’s output logits, we optimize the following distillation loss:
\begin{equation}
\mathcal{L}_{\text{distill}} = \frac{1}{d} \sum_{i=1}^d (z_{\text{PUF},i} - z_{s,i})^2
\end{equation}
where $d$ is the number of the classes. The loss function minimizes the difference between the student’s outputs and the fingerprinted teacher outputs, ensuring that the student inherits both the model’s knowledge and the embedded fingerprint.

By superimposing PUF-based logit fingerprinting directly onto the teacher’s output and transferring them through knowledge distillation, our method achieves device-level traceability. Even if the teacher is stolen and new students are cloned from it, the behavioral fingerprint remains detectable, allowing the leaked model to be traced back to the device.

\subsection{Simulating Realistic PUF Noisy Behavior}

While our implementation is simulation-based, it is designed to reflect the noisy behavior typically seen in real PUF hardware. For each experiment, we assign a unique PUF key to represent the identity of a \emph{leaker device}. This selected PUF key $\mathbf{k} \in \{0,1\}^n$ is duplicated in all samples in a batch to form a key matrix $\mathbf{K} \in \{0,1\}^{B \times n}$, where $B$ is the batch size and each row of $\mathbf{K}$ is identical. To mimic intra-device variation observed in real-world PUF behavior, we generate a binary noise mask $\mathbf{M} \in \{0,1\}^{B \times n}$, where each element $M_{i,j}$ is sampled independently from a Bernoulli distribution with a fixed bit error rate $p_{\text{flip}}$:
\begin{equation}
M_{i,j} \sim \text{Bernoulli}(p_{\text{flip}})
\end{equation}
We then apply bitwise XOR between the clean key matrix and the noise mask to produce a noisy version of the PUF key matrix $\tilde{\mathbf{K}} $ by using $\tilde{\mathbf{K}} = \mathbf{K} \oplus \mathbf{M}$. The perturbated key matrix $\tilde{\mathbf{K}}$ is then used to compute logit perturbations using Equation (\ref{eq:delta_definition}).
These perturbations are finally added to the clean teacher logits to obtain fingerprinted outputs. This process ensures that the embedded fingerprint reflects hardware-consistent noise patterns, thereby enhancing the realism and robustness of our simulation.

\subsection{Two-Stage signature recovery process}
\label{sec:decoder}

In this work, we introduce a \textbf{two-stage signature recovery process} including neural network-based decoder and Hamming distance decoder as shown in Figure~\ref{fig:system}  for recovering the PUF key from the student logits. The recovery process leverages the structured nature of PUF perturbations integrated into the teacher's logits and the consistent patterns learned by the student. 

\subsubsection{Neural Network-based Decoder}

The first stage decoder involves training a neural network to recover PUF keys from the outputs of a student model, as shown in Algorithm~\ref{alg:decoder_train}. Since real student models corresponding to every possible PUF key are not available, we construct a synthetic dataset that simulates the fingerprinting and learning process under controlled conditions. This approach is valid because the decoder does not rely on the internal parameters of any specific student model; rather, it depends on the statistical structure of the embedded fingerprint and noise, which can be reproduced in simulation. By modeling perturbations, teacher variability, and student noise, the synthetic dataset captures the essential patterns and enables robust training that generalizes to real-world deployment.

We sample $R$ distinct keys from the key space of $2^n$ to represent simulated devices. For each device $i$, we generate $Q$ logit samples under random perturbation scales $\epsilon \in \mathcal{E}$ and noise levels $\sigma$, reflecting variations in real systems. Teacher logits $\mathbf{z}_t^{(j)} \sim \mathcal{N}(\mathbf{0}, \mathbf{I}_d)$ are perturbed by a perturbation vector $\boldsymbol{\delta}_i$ for the device $\mathbf{k}_i$ via Equation~(\ref{eq:delta_definition}), and Gaussian noise $\boldsymbol{\eta}$ is injected to simulate student behavior:
\begin{equation}
\mathbf{z}_{\text{s}}^{(j)} = \mathbf{z}_t^{(j)} + \boldsymbol{\delta}_i + \boldsymbol{\eta}, \quad \boldsymbol{\eta} \sim \mathcal{N}(\mathbf{0}, \sigma^2 \mathbf{I}_d).    
\end{equation}
where $j$ indexes the $Q$ independently sampled teacher logits per device. The logit difference $\Delta \mathbf{z}^{(j)} = \mathbf{z}_{\text{s}}^{(j)} - \mathbf{z}_t^{(j)}$ encodes both the embedded PUF fingerprint and stochastic variations. Each difference vector is paired with the corresponding key $\mathbf{k}_i$, forming labeled training pairs $(\Delta \mathbf{z}^{(j)}, \mathbf{k}_i)$.  

After constructing all $R \times Q$ samples, we train a multi-layer perceptron decoder $\mathcal{D}_\theta$ to recover the PUF key from each logit difference vector $\Delta \mathbf{z}^{(j)}$. It outputs a probability vector:
\[
\mathbf{p}_k^{(j)} = \mathcal{D}_\theta(\Delta \mathbf{z}^{(j)}),
\]
The model is trained using binary cross-entropy loss between the predicted probabilities and the ground-truth key:
\[
\mathcal{L}_{\text{BCE}}^{(j)} = - \frac{1}{n} \sum_{b=1}^{n} \left[
k_{i,b} \log p_{k_{i,b}}^{(j)} + (1 - k_{i,b}) \log (1 - p_{k_{i,b}}^{(j)})
\right].
\]
where each element $p_{k_{i,b}}^{(j)}$ represents the predicted probability that bit $b$ of the key $\mathbf{k}_i$ is 1.
At inference time, each bit is predicted as 1 if the probability is greater than 0.5, and 0 otherwise. Performance of the decoder is evaluated by bitwise accuracy.

This synthetic training pipeline provides an efficient and scalable way to achieve robust behavior-to-identity mapping, while avoiding costly real-world data collection and ensuring strong generalization across diverse deployment scenarios.

\begin{algorithm}
\caption{PUF Logit Simulation and Decoder Training with Multiple Perturbation Scales}
\label{alg:decoder_train}
\begin{algorithmic}[1]
\State \textbf{Input:} Bit-length $n$, number of devices $R$, samples per device $Q$, perturbation set $\mathcal{E}$, student noise level $\sigma$
\State \textbf{Output:} Trained decoder $\mathcal{D}_\theta$
\State Generate $R$ unique keys from $\{0,1\}^n$ to form subset $\mathcal{K}$
\State Initialize training dataset $(\mathcal{X}, \mathcal{Y})$ 
\For{each device key $\mathbf{k}_i$ in $\mathcal{K}$}
    \State Randomly sample perturbation scale $\epsilon \sim \mathcal{E}$
    \For{$j = 1$ to $Q$}
        \State Generate clean teacher logits $\mathbf{z}_t^{(j)} \sim \mathcal{N}(0, I_d)$
        \State Compute perturbation vector $\boldsymbol{\delta}_i = \epsilon \cdot (1 - 2 \cdot \mathbf{k}_i)$
        \State Perturb teacher logits: $\mathbf{z}_{\text{PUF}}^{(j)} \gets \mathbf{z}_t^{(j)} + \boldsymbol{\delta}_i$
        \State Simulate student logits: $\mathbf{z}_{\text{s}}^{(j)} \gets \mathbf{z}_{\text{PUF}}^{(j)} + \boldsymbol{\eta}$
        \State Compute logit difference: $\Delta \mathbf{z}^{(j)} = \mathbf{z}_{\text{s}}^{(j)} - \mathbf{z}_t^{(j)}$
        \State Append training pair $(\Delta \mathbf{z}^{(j)}, \mathbf{k}_i)$ to $(\mathcal{X}, \mathcal{Y})$
    \EndFor
\EndFor
\State Train decoder $\mathcal{D}_\theta$ on $(\mathcal{X}, \mathcal{Y})$ using binary cross-entropy loss with sigmoid activation and early stopping
\State \Return trained decoder $\mathcal{D}_\theta$
\end{algorithmic}
\end{algorithm}

\subsubsection{Hamming Distance Decoder}

While the neural network decoder is effective at recovering the original PUF key, minor prediction errors may occur due to noise or decoder uncertainty. To address this, we introduce a second-stage refinement step based on a Hamming distance decoder. In this stage, the predicted binary key $\hat{\mathbf{k}}$ output from the neural decoder is compared against a predefined database $\mathcal{K}$ of registered PUF keys. The final recovered key $\mathbf{k}^*$ is selected as the entry in $\mathcal{K}$ that has the minimum Hamming distance to the predicted key by using $\mathbf{k}^* = \arg\min_{\mathbf{k} \in \mathcal{K}} d_H(\hat{\mathbf{k}}, \mathbf{k})$, where \( d_H(\cdot, \cdot) \) denotes the Hamming distance. This refinement step is a lightweight error correction mechanism to correct small bit errors. The performance of the two-stage decoder has been evaluated in Section~\ref{sec:experiment}.
By combining a neural network decoder with a lightweight Hamming distance-based step, the method offers robust predictions. 

\begin{table}[h]
\centering
\caption{$m$-bit perturbation mapping ($m=3$, $\epsilon = 0.4$)}
\label{tab:bitmap-general}
\begin{tabular}{c|c|c|c}
\hline
\hline
Bits $(b_0 b_1 b_2)$ & Integer $U$ & Shifted $(U + 0.5)$ & Perturbation $\delta$ \\
\hline
000 & 0   & +0.5  & +0.2 \\
001 & +1  & +1.5  & +0.6 \\
010 & +2  & +2.5  & +1.0 \\
011 & +3  & +3.5  & +1.4 \\
111 & -1  & -0.5  & -0.2 \\
110 & -2  & -1.5  & -0.6 \\
101 & -3  & -2.5  & -1.0 \\
100 & -4  & -3.5  & -1.4 \\
\hline
\hline
\end{tabular}
\end{table}

\subsection{Bit Compression Scheme}
\label{sec:bitmapped}

Mapping one PUF bit to each output logit limits scalability: with $n$ logits, at most $2^n$ unique devices can be supported, and the effective key space is even smaller due to bias and instability. This makes one-bit-per-logit encoding unsuitable for deployments that require millions or billions of identifiers.

To overcome this limitation, we introduce a \textbf{bit compression} strategy to map multiple PUF keys to each logit. The PUF key is partitioned into fixed-length segments of $m$ bits, with each segment assigned to a single logit. In this formulation, each logit can represent $2^m$ distinct values, thereby expanding the identity space exponentially without increasing the dimensionality of the model’s output. The benefits are obvious. With 10 logits, encoding one bit per logit supports $2^{10} = 1{,}024$ unique identities. Increasing the density to two bits per logit increases the capacity to $2^{20} \approx 10^6$ identities, while three bits per logit extends the capacity further to $2^{30} \approx 10^9$ identities. This exponential growth enables large-scale deployment while preserving a compact model architecture.

To implement this encoding, each $m$-bit segment $(b_0 b_1 \dots b_{m-1})$ is interpreted as a signed integer $U \in \{-2^{m-1}, \dots, 2^{m-1} - 1\}$ using standard two's complement representation. The integer value is then linearly mapped to a logit perturbation as $\delta = \epsilon \cdot (U + 0.5)$, where $\epsilon$ is a hyperparameter that controls perturbation levels. The $+0.5$ shift ensures the perturbation are symmetrically distributed around zero. Table~\ref{tab:bitmap-general} shows an example for $m=3$ and $\epsilon=0.4$.

\begin{table*}[t]
\centering
\small
\caption{Baseline model architectures and teacher accuracies across datasets (without PUF perturbation)}
\label{tab:model-arch-acc}
\begin{tabular}{l|l|l|c}
\hline
\hline
\textbf{Dataset}  & \textbf{Teacher Architecture}& \textbf{Student Architecture} & \textbf{Teacher Accuracy}   \\
\hline
CIFAR-10 & 64c3-64c3-2s-128c3-128c3-2s-F128-F10 & 64c3-2s-128c3-2s-128c3-2s-F128-F10 & 79.36\% \\
CIFAR-20 & 64c3-64c3-2s-128c3-128c3-2s-F128-F20 & 64c3-2s-128c3-2s-128c3-2s-F128-F20 & 64.85\% \\
CIFAR-50 & 64c3-64c3-2s-128c3-128c3-2s-256c3-256c3-2s-GAP-F256-F50 & 64c3-2s-128c3-2s-128c3-2s-F256-F50 & 66.62\% \\
\hline
\hline
\end{tabular}
\vspace{-3mm}
\end{table*}

\section{Experiments}

\subsection{Experimental Setup}

To evaluate the effectiveness of the proposed PUF-based fingerprinting framework, 
we assess both classification performance and the recoverability of PUF keys. 
Recoverability is measured using bit error rate (BER), the fraction of incorrect 
bits in reconstructed keys, and frame error rate (FER), the fraction of keys 
containing at least one error. We evaluate our method on three image classification 
datasets: \textbf{CIFAR-10}, \textbf{CIFAR-20}, and \textbf{CIFAR-50}. 
Table~\ref{tab:model-arch-acc} summarizes the teacher–student architectures 
and their baseline accuracies. For CIFAR-10 and CIFAR-20, a convolutional teacher 
achieves 79.36\% and 64.85\%, respectively, while the deeper CIFAR-50 teacher 
with additional convolutional and global-average-pooling layers reaches 66.62\%.

We adopt the knowledge distillation setup illustrated in Figure~\ref{fig:system}. 
To emulate real deployment, we simulate PUF responses by generating a pool of 
binary keys, selecting several devices, and designating one as the leaker. 
A 5\% bit-flip rate is applied to model PUF noise, and the resulting noisy key 
is mapped into a perturbation vector that is added to the teacher’s clean logits 
before distillation.

\begin{table}[h]
\centering
\small
\caption{Accuracy–BER/FER(\%) trade-off under varying perturbation levels $\epsilon$}
\label{tab:epsilon-tradeoff}
\begin{tabular}{l|l|c|c|c|c}
\hline
\hline
\textbf{Model} & \textbf{$Acc_s$} & \textbf{$\epsilon$} & \textbf{$Acc_p$} & \textbf{BER} & \textbf{FER}\\
\hline
CIFAR-10 &82.4 & 0.01 & $81.63 \pm 0.39$ &24 & 55  \\
         & & 0.02 & $81.67 \pm 0.55$ &6  &15\\
        &  & 0.05 & $81.16 \pm 0.56$  &0 & 0 \\
\hline
CIFAR-20 &64.2 & 0.01 & $64.41 \pm 0.71$ &16.7&44  \\
         & & 0.02 & $64.35 \pm 0.93$ & 0.45&2\\
         & & 0.05 & $62.03 \pm 1.74$ &0  &0\\
\hline
CIFAR-50 & 52.4 & 0.02 & $51.92 \pm 1.40$ &15.14& 34\\
        &  & 0.05 &$52.44 \pm 1.53$  & 0&0\\
       &   & 0.1 & $50.97 \pm 9.25$ &0&0\\
\hline
\hline
\end{tabular}

\vspace{0.4em}
\footnotesize{\textit{Note.} $Acc_s$ denotes the baseline student accuracy obtained from standard knowledge distillation without PUF perturbation. 
$Acc_p$ represents the student accuracy when PUF-derived perturbations are added to teacher logits with varying $\epsilon$ values. 
BER and FER denote the bit error rate and frame error rate of fingerprint recovery, respectively.}

\vspace{-5mm}

\end{table}

\subsection{Tradeoff Between Student Accuracy and Key Recoverability}  
\label{sec:experiment}
We evaluate how the perturbation level $\epsilon$ affects both \emph{student accuracy} and \emph{PUF recovery rate}. Each configuration is averaged over 100 trials with different leaker devices. Table~\ref{tab:epsilon-tradeoff} summarizes these results, where $Acc_s$ denotes the clean student accuracy without PUF perturbation and $Acc_p$ denotes the accuracy under perturbation.  

Across all datasets, a consistent trend is observed: increasing the perturbation strength enhances fingerprint recovery, often achieving perfect reconstruction (BER = 0, FER = 0) at moderate-to-high $\epsilon$ values, while student accuracy remains close to the baseline. For CIFAR-10 and CIFAR-20, small perturbations ($\epsilon=0.01$) leads to weak recovery (BER/FER = $24\%/55\%$ and $16.7\%/44\%$, respectively), but increasing $\epsilon$ to 0.05 enables error-free decoding with higher accuracy variance. CIFAR-50 has similar behavior, but requires larger perturbation to achieve complete recovery: $\epsilon=0.05$ and above ensure perfect recovery but introduces higher accuracy variance ($\pm 9.25$). In some cases, students trained with larger perturbations achieve slightly higher accuracy than those with smaller $\epsilon$, as stronger noise can introduce more signal diversity and helps prevent overfitting \cite{noise}. We also observe that some student models even outperform the teacher without perturbation, showing the known benefit of knowledge distillation in improving generalization through soft supervision.  

\begin{figure}[htbp]
\centering
\makebox[\textwidth][l]{
\hspace*{-0.2em}
\begin{minipage}{0.4\textwidth}
\centering
\begin{tikzpicture}
\begin{axis}[
    ybar,
    bar width=5pt,
    width=\textwidth,
    height=0.6\linewidth,
    scale only axis,
    xmin=0, xmax=3.5,
    ymin=0, ymax=50,
    ylabel={\%},
    xtick={0.5, 2, 3.5},
    xticklabels={CIFAR-10, CIFAR-20, CIFAR-50},
    xticklabel style={font=\scriptsize, yshift=-5pt},
    ylabel style={font=\scriptsize},
    tick label style={font=\scriptsize},
    nodes near coords,
    every node near coord/.append style={font=\tiny},
    legend style={font=\tiny, at={(0.5,-0.25)}, anchor=north, legend columns=2},
    axis lines=box,
    clip=false,
    enlarge x limits=0.2,
    ymajorgrids=true,
    xmajorgrids=true,
    grid style=dashed,
    ytick={10,20,30,40,50},
    extra x tick style={grid style=dashed},
]

\pgfmathsetmacro{\shift}{0.06}


\addplot+[draw=olive!80!black, fill=olive!30] coordinates {(1 - 0.3*\shift, 32.0)};
\addplot+[draw=olive!80!black, fill=olive!30] coordinates {(1 + 0.3*\shift, 13.5)};
\addplot+[draw=olive!80!black, fill=olive!30] coordinates {(1 + 0.5*\shift, 2.8)};
\node[font=\tiny, anchor=north east, rotate=45] at (axis cs:1 - 12*\shift, 1) {0.01};
\node[font=\tiny,anchor=north east, rotate=45] at (axis cs:1 - 9*\shift, 1) {0.02};
\node[font=\tiny, anchor=north east, rotate=45] at (axis cs:1 - 6*\shift, 1) {0.05};

\addplot+[draw=orange!80!black, fill=orange!40] coordinates {(2 - 0.3*\shift, 22.5)};
\addplot+[draw=orange!80!black, fill=orange!40] coordinates {(2 + 0.3*\shift, 8.1)};
\addplot+[draw=orange!80!black, fill=orange!40] coordinates {(2 + 0.5*\shift, 0.0)};
\node[font=\tiny, anchor=north east, rotate=45] at (axis cs:2 -5*\shift, 1) {0.01};
\node[font=\tiny, anchor=north east, rotate=45] at (axis cs:2 - 1*\shift, 1) {0.02};
\node[font=\tiny, anchor=north east, rotate=45] at (axis cs:2 + 2*\shift, 1) {0.05};

\addplot+[draw=teal!80!black, fill=teal!30] coordinates {(3 - 0.5*\shift, 40.3)};
\addplot+[draw=teal!80!black, fill=teal!30] coordinates {(3 - 0.3*\shift, 29.5)};
\addplot+[draw=teal!80!black, fill=teal!30] coordinates {(3 + 0.3*\shift, 8.44)};
\addplot+[draw=teal!80!black, fill=teal!30] coordinates {(3 + 0.5*\shift, 0.8)};
\node[font=\tiny, anchor=north east, rotate=45] at (axis cs:3 + 3*\shift, 1) {0.01};
\node[font=\tiny, anchor=north east, rotate=45] at (axis cs:3 + 6*\shift, 1) {0.02};
\node[font=\tiny, anchor=north east, rotate=45] at (axis cs:3 + 10*\shift, 1) {0.05};
\node[font=\tiny, anchor=north east, rotate=45] at (axis cs:3 + 13*\shift, 1) {0.1};
\end{axis}
\end{tikzpicture}
\end{minipage}
}
\captionsetup{justification=raggedright, singlelinecheck=false, margin={0.5cm,0cm}}
\caption{PUF BER after first decoder across different PUF lengths and perturbation levels.}
\label{fig:BER_accuracy-recovery-tradeoff}
\end{figure}
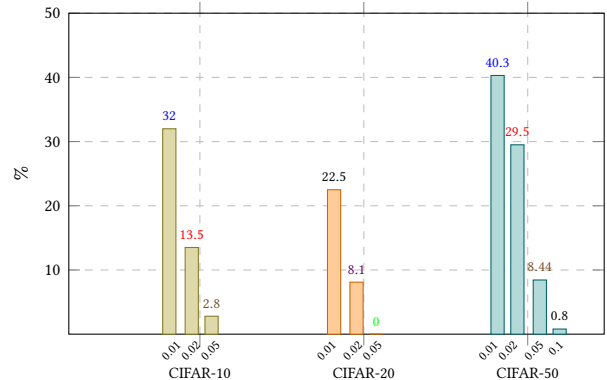

Figure~\ref{fig:BER_accuracy-recovery-tradeoff} reports the BER after using the neural network decoder with different perturbation levels and datasets. As expected, higher perturbation strengths $\epsilon$ generally reduce BER. For CIFAR-10, BER decreases from 32\% at $\epsilon=0.01$ to below 3\% at $\epsilon=0.05$. A similar trend is observed for CIFAR-20, where BER drops from 22.5\% to nearly zero as $\epsilon$ increases. In CIFAR-50, which presents a more challenging recovery setting due to its larger output space, BER declines from 29.5\% at $\epsilon=0.02$ to below 1\% at $\epsilon=0.10$. These results show that small perturbations lead to weaker and noisier recovery, while moderate perturbation levels are sufficient to achieve consistently low BER across datasets.

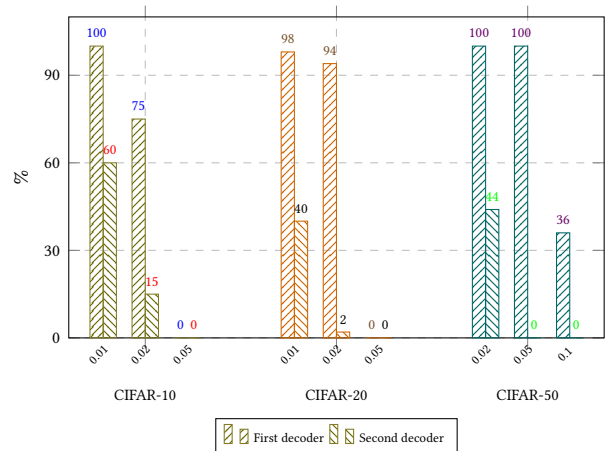
\begin{figure}[htbp]
\centering
\makebox[\textwidth][l]{
\hspace*{-1.0em}
\begin{minipage}{0.4\textwidth}
\centering
\begin{tikzpicture}
\begin{axis}[
    ybar,
    bar width=5pt,
    width=\textwidth,
    height=0.6\linewidth,
    scale only axis,
    xmin=0.6, xmax=3.4,
    ymin=0, ymax=110,
    ylabel={\%},
    xtick={1,2,3},
    xticklabels={CIFAR-10, CIFAR-20, CIFAR-50},
    xticklabel style={font=\scriptsize, yshift=-12pt},
    ylabel style={font=\scriptsize},
    tick label style={font=\scriptsize},
    nodes near coords,
    every node near coord/.append style={font=\tiny},
    legend style={font=\tiny, at={(0.5,-0.25)}, anchor=north, legend columns=2},
    axis lines=box,
    clip=false,
    enlarge x limits=0,
    ymajorgrids=true,
    xmajorgrids=true,
    grid style=dashed,
    ytick={0,30,60,90},
]

\pgfmathsetmacro{\sub}{0.22}
\newcommand{\pairshift}{2.5pt}

\def\AtenOne{100.0}   \def\BtenOne{60.0}
\def\AtenTwo{75.0}   \def\BtenTwo{15.0}
\def\AtenFive{0.0}   \def\BtenFive{0.0}
\def\AtwentyOne{98.0}  \def\BtwentyOne{40.0}
\def\AtwentyTwo{94.0}   \def\BtwentyTwo{2.0}
\def\AtwentyFive{0.0}  \def\BtwentyFive{0.0}
\def\AfiftyOne{100.0}   \def\BfiftyOne{44.0}
\def\AfiftyTwo{100.0}   \def\BfiftyTwo{0.0}
\def\AfiftyFive{36.0}   \def\BfiftyFive{0.0}

\addplot+[draw=olive!80!black, fill=olive!30,
          pattern=north east lines, pattern color=olive!80!black,
          bar shift=-\pairshift]
  coordinates { (1-\sub,\AtenOne) (1,\AtenTwo) (1+\sub,\AtenFive) };
\addplot+[draw=olive!80!black, fill=olive!10,
          pattern=north west lines, pattern color=olive!80!black,
          bar shift=\pairshift]
  coordinates { (1-\sub,\BtenOne) (1,\BtenTwo) (1+\sub,\BtenFive) };

\addplot+[draw=orange!80!black, fill=orange!40,
          pattern=north east lines, pattern color=orange!80!black,
          bar shift=-\pairshift]
  coordinates { (2-\sub,\AtwentyOne) (2,\AtwentyTwo) (2+\sub,\AtwentyFive) };
\addplot+[draw=orange!80!black, fill=orange!15,
          pattern=north west lines, pattern color=orange!80!black,
          bar shift=\pairshift]
  coordinates { (2-\sub,\BtwentyOne) (2,\BtwentyTwo) (2+\sub,\BtwentyFive) };

\addplot+[draw=teal!80!black, fill=teal!30,
          pattern=north east lines, pattern color=teal!80!black,
          bar shift=-\pairshift]
  coordinates { (3-\sub,\AfiftyOne) (3,\AfiftyTwo) (3+\sub,\AfiftyFive) };
\addplot+[draw=teal!80!black, fill=teal!10,
          pattern=north west lines, pattern color=teal!80!black,
          bar shift=\pairshift]
  coordinates { (3-\sub,\BfiftyOne) (3,\BfiftyTwo) (3+\sub,\BfiftyFive) };

\addlegendentry{First decoder}
\addlegendentry{Second decoder}

\node[font=\tiny, anchor=north, rotate=45, yshift=-4pt] at (axis cs:1-1.5*\sub,0) {0.01};
\node[font=\tiny, anchor=north, rotate=45, yshift=-4pt] at (axis cs:1-0.5*\sub,     0) {0.02};
\node[font=\tiny, anchor=north, rotate=45, yshift=-4pt] at (axis cs:1+0.5*\sub,0) {0.05};

\node[font=\tiny, anchor=north, rotate=45, yshift=-4pt] at (axis cs:2-1.5*\sub,0) {0.01};
\node[font=\tiny, anchor=north, rotate=45, yshift=-4pt] at (axis cs:2-0.5*\sub,     0) {0.02};
\node[font=\tiny, anchor=north, rotate=45, yshift=-4pt] at (axis cs:2+0.5*\sub,0) {0.05};

\node[font=\tiny, anchor=north, rotate=45, yshift=-4pt] at (axis cs:3-1.5*\sub,0) {0.02};
\node[font=\tiny, anchor=north, rotate=45, yshift=-4pt] at (axis cs:3-0.5*\sub,     0) {0.05};
\node[font=\tiny, anchor=north, rotate=45, yshift=-4pt] at (axis cs:3+0.5*\sub,0) {0.1};

\end{axis}
\end{tikzpicture}
\end{minipage}
}
\caption{PUF FER after first decoder and second decoder across different PUF lengths and perturbation levels.}
\label{fig:FER_accuracy-recovery-tradeoff}
\vspace{-3mm}
\end{figure}

Figure~\ref{fig:FER_accuracy-recovery-tradeoff} highlights the role of the Hamming-distance decoder, which significantly improves FER. Although the neural network decoder alone reduces BER, minor key errors may still occur, especially under low perturbation strengths. The second stage reliably  corrects these  errors. For example, in CIFAR-10 with $\epsilon=0.02$, the FER after the first decoder is 75\%, but the Hamming-distance decoder reduces this to 15\%. Similar corrections occur for CIFAR-20 and CIFAR-50, where the second decoder achieves perfect recovery under conditions where the first decoder fails. These results confirm the effectiveness of the two-stage decoding process: the neural decoder captures the PUF signal, and the second stage enhances reliability by correcting residual bit errors.


\subsection{Simulation for Bit Compression scheme}

\begin{figure}[htbp]
\centering
\makebox[\textwidth][l]{
\hspace*{-1.5em}
\begin{minipage}{0.4\textwidth}
\centering
\begin{tikzpicture}
\begin{axis}[
    ybar,
    bar width=5pt,
    width=\textwidth,
    height=0.6\linewidth,
    scale only axis,
    xmin=0, xmax=3.7,
    ymin=0, ymax=110,
    ylabel={\%},
    xtick={0.3, 2, 3.7},
    xticklabels={10-bit, 20-bit, 30-bit},
    xticklabel style={font=\scriptsize, yshift=-5pt},
    ylabel style={font=\scriptsize},
    tick label style={font=\scriptsize},
    every node near coord/.append style={font=\tiny},
    legend style={font=\tiny, at={(0.5,-0.18)}, anchor=north, legend columns=2},
    axis lines=box,
    clip=false,
    enlarge x limits=0.2,
    ymajorgrids=true,
    xmajorgrids=true,
    grid style=dashed,
    ytick={20,40,60,80,100},
    extra x tick style={grid style=dashed},
]

\pgfmathsetmacro{\shift}{0.06}

\addplot+[draw=blue, fill=blue!30, error bars/.cd, y dir=both, y explicit] 
coordinates {(1 - 0.5*\shift, 81.67) +- (0, 0.55)};
\node[font=\tiny, anchor=east, xshift=-32pt, yshift=4pt ] 
    at (axis cs:1 - 0.5*\shift, 81.67) {81.67 $\pm$ 0.55};

\addplot+[draw=red, fill=red!30] 
coordinates {(1 - 0.5*\shift, 15.0)};
\node[font=\tiny, anchor=south, xshift=-32pt] 
    at (axis cs:1 - 0.5*\shift, 15.0) {15};

\addplot+[draw=blue, fill=blue!30, error bars/.cd, y dir=both, y explicit] 
coordinates {(1 - 0.1*\shift, 81.16) +- (0, 0.56)};
\node[font=\tiny, anchor=east, xshift=2pt, yshift=4pt ] 
    at (axis cs:1 - 0.1*\shift, 81.67) {81.67 $\pm$ 0.56};
\addplot+[draw=red, fill=red!30] 
coordinates {(1 - 0.3*\shift, 0.0)};
\node[font=\tiny, anchor=south, xshift=-17pt] 
    at (axis cs:1 - 0.3*\shift, 0.0) {0};
    

\node[font=\tiny, anchor=north east, rotate=45] at (axis cs:0.1, 2) {0.02};
\node[font=\tiny,anchor=north east, rotate=45] at (axis cs:0.5, 2) {0.05};

\addplot+[draw=blue, fill=blue!30, error bars/.cd, y dir=both, y explicit] 
coordinates {(2 - 0.5*\shift, 65.99) +- (0, 8.31)};
\node[font=\tiny, anchor=east, xshift=-2pt, yshift=15pt ] 
    at (axis cs:2 - 0.5*\shift, 65.99) {65.99 $\pm$ 8.31};

\addplot+[draw=red, fill=red!30] 
coordinates {(2 - 0.3*\shift, 32.0)};
\node[font=\tiny, anchor=south, xshift=-4pt] 
    at (axis cs:2 - 0.3*\shift, 32.0) {32};

\addplot+[draw=blue, fill=blue!30, error bars/.cd, y dir=both, y explicit] 
coordinates {(2 - 0.1*\shift, 70.22) +- (0, 10.19)};
\node[font=\tiny, anchor=east, xshift=25pt, yshift=20pt ] 
    at (axis cs:2 - 0.1*\shift, 70.22) {70.22 $\pm$ 10.19};

\addplot+[draw=red, fill=red!30] 
coordinates {(2 + 0.3*\shift, 0.0)};
\node[font=\tiny, anchor=south, xshift=10pt] 
    at (axis cs:2 + 0.3*\shift, 0.0) {0};
\node[font=\tiny, anchor=north east, rotate=45] at (axis cs:1.8, 2) {0.1};
\node[font=\tiny, anchor=north east, rotate=45] at (axis cs:2.2, 2) {0.2};
\addplot+[draw=blue, fill=blue!30, error bars/.cd, y dir=both, y explicit] 
coordinates {(3 - 0.5*\shift, 83.22) +- (0, 15.65)};
\node[font=\tiny, anchor=east, xshift=20pt, yshift=25pt ] 
    at (axis cs:3 - 0.5*\shift, 83.22) {83.22 $\pm$ 15.65};
\addplot+[draw=red, fill=red!30] 
coordinates {(3 - 0.3*\shift, 57.0)};
\node[font=\tiny, anchor=south, xshift=25pt] 
    at (axis cs:3 - 0.3*\shift, 57.0) {57};

\addplot+[draw=blue, fill=blue!30, error bars/.cd, y dir=both, y explicit] 
coordinates {(3 - 0.1*\shift, 82.82) +- (0, 16.54)};
\node[font=\tiny, anchor=east, xshift=61pt, yshift=25pt ] 
    at (axis cs:3 - 0.1*\shift, 82.82) {82.82 $\pm$ 16.54};
    
\addplot+[draw=red, fill=red!30] 
coordinates {(3 + 0.3*\shift, 0.0)};
\node[font=\tiny, anchor=south, xshift=37pt] 
    at (axis cs:3 + 0.3*\shift, 0.0) {0};
\node[font=\tiny, anchor=north east, rotate=45] at (axis cs:3.5, 2) {0.3};
\node[font=\tiny, anchor=north east, rotate=45] at (axis cs:3.9, 2) {0.4};

\legend{Student Accuracy, PUF FER}
\end{axis}
\end{tikzpicture}
\end{minipage}
}
\caption{Trade-off between student accuracy (CIFAR-10) and PUF FER for Bit Compression scheme}
\label{fig:compressed-accuracy-recovery-tradeoff}
\end{figure}
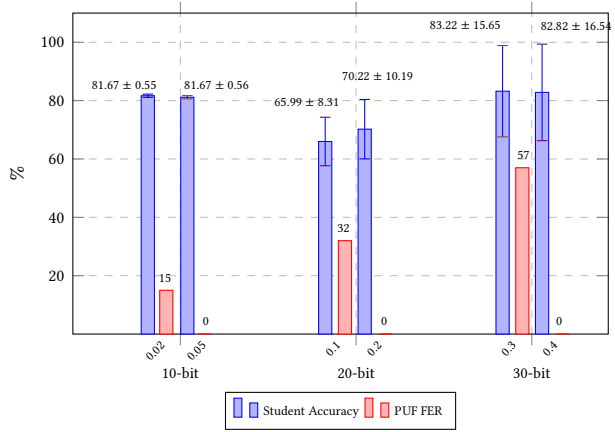

We evaluated the scalability of the bit-compressed fingerprinting framework by mapping longer PUF keys into a fixed logit vector, as described in Section~\ref{sec:bitmapped}. This setup reflects practical scenarios where a large identity space must be supported despite limited output dimensionality.
We tested PUF lengths of 10, 20, and 30 bits. For each case, PUF keys were generated, compressed, and superimposed onto teacher logits, and student models were trained under different perturbation levels~$\epsilon$. Figure~\ref{fig:compressed-accuracy-recovery-tradeoff} summarizes the resulting trade-offs. With 10-bit keys, student accuracy remains stable above 80\% and recovery is moderate at small~$\epsilon$. Increasing the key length to 20 bits expands identity capacity but introduces greater accuracy variation (57--80\%) and requires higher~$\epsilon$ for reliable recovery. The 30-bit setting further enlarges the identity space, but recovery becomes highly sensitive to~$\epsilon$ and accuracy variation increases, although accuracy stays above 60\%.

Overall, longer PUF keys offer exponentially larger capacity but amplify the accuracy–recovery trade-off. Proper tuning of the perturbation level~$\epsilon$ is therefore critical for scalable deployment.

\subsection{Comparison with Existing schemes}

Table~\ref{tab:comparison} summarizes several representative
fingerprinting and protection schemes and compares them along
four key aspects: traceability, internal model access,  
hardware awareness (HW-Aware), and  Overhead.
Traceability indicates whether the source can be identified 
when a model is leaked. Internal access denotes whether the scheme
requires accessing model parameters. Hardware awareness captures whether the
method leverages device-specific properties. We also report the approximate overhead of
each scheme. 

Traditional watermarking methods~\cite{Uchida_2017,USENIX} aim to prove ownership but often require internal model access, making them vulnerable to removal by fine-tuning. Hardware-based defenses~\cite{ISCAS,privacypreserving} strengthen extraction resistance but primarily focus on theft prevention and introduce additional overhead.
In contrast, PUF-Logit fingerprinting addresses a complementary goal: post-theft traceability. Instead of stopping theft, it identifies the device from which a stolen or redistributed model originated. Because it works at the logit level and requires no architectural changes, PUF-Logit is lightweight, hardware-aware, and easily combined with existing defenses to provide a layered protection strategy.

\begin{table}[t]
\centering
\caption{Comparison of existing protection approaches}
\resizebox{\columnwidth}{!}{
\begin{tabular}{lcccc}
\hline
\textbf{Method} & \textbf{Traceable} & \textbf{No Internal Access} &
\textbf{HW-Aware} & \textbf{Overhead} \\
\hline

~\cite{USENIX}         
& \textcolor{red}{\ding{55}}    %
& \textcolor{red}{\ding{55}}    %
& \textcolor{red}{\ding{55}}    %
& Med \\

~\cite{privacypreserving} 
& \textcolor{red}{\ding{55}}
& \textcolor{green}{\ding{51}}  %
& \textcolor{green}{\ding{51}}
& High \\

~\cite{ISCAS}           
& \textcolor{red}{\ding{55}}
& \textcolor{green}{\ding{51}}
& \textcolor{green}{\ding{51}}
& High \\

~\cite{Uchida_2017} 
& \textcolor{red}{\ding{55}}
& \textcolor{red}{\ding{55}}
& \textcolor{red}{\ding{55}}
& Low \\

\textbf{ours}     
& \textcolor{green}{\textbf{\ding{51}}}
& \textcolor{green}{\textbf{\ding{51}}}
& \textcolor{green}{\textbf{\ding{51}}}
& \textbf{Low} \\

\hline
\end{tabular}
}
\label{tab:comparison}
\end{table}








\section{Conclusion and Discussion}

In this work, we presented a lightweight, hardware-rooted framework that embeds device-specific PUF fingerprints into neural networks through knowledge distillation, enabling post-theft traceability  with no architectural changes or heavy runtime mechanisms. By perturbing teacher logits with PUF-derived signals, the fingerprints naturally propagate into student models while maintaining negligible accuracy degradation, making the approach suitable for resource-constrained deployment. Experiments on CIFAR-10, CIFAR-20, and CIFAR-50 demonstrate reliable fingerprint embedding and recovery under realistic PUF noise. The two-stage decoder improves recovery robustness, and the bit-compression scheme expands fingerprinting capacity, establishing PUF-based logit perturbation as an efficient solution for tracing stolen models in hardware settings.

Although our evaluation uses simulated PUF data, we incorporate realistic hardware effects such as bit-flip noise and device-level variation. As future work, we plan to deploy PUF-Logit on FPGA hardware to validate fingerprint stability and key recovery under real PUF conditions.
While our framework is designed for the distillation-based theft scenario, the idea can extend to broader threats. Combining PUF-guided perturbations with runtime monitoring or side-channel–aware defenses could offer more comprehensive protection beyond distillation-based attacks.


\clearpage
\balance

\bibliographystyle{ACM-Reference-Format}
\bibliography{refs}

@inproceedings{Uchida_2017, series={ICMR ’17},
   title={Embedding Watermarks into Deep Neural Networks},
   url={http://dx.doi.org/10.1145/3078971.3078974},
   DOI={10.1145/3078971.3078974},
   booktitle={Proceedings of the 2017 ACM on International Conference on Multimedia Retrieval},
   publisher={ACM},
   author={Uchida, Yusuke and Nagai, Yuki and Sakazawa, Shigeyuki and Satoh, Shin’ichi},
   year={2017},
   month=jun, pages={269–277},
   collection={ICMR ’17} }

@inproceedings {USENIX,
author = {Yossi Adi and Carsten Baum and Moustapha Cisse and Benny Pinkas and Joseph Keshet},
title = {Turning Your Weakness Into a Strength: Watermarking Deep Neural Networks by Backdooring},
booktitle = {27th USENIX Security Symposium (USENIX Security 18)},
year = {2018},
isbn = {978-1-939133-04-5},
address = {Baltimore, MD},
pages = {1615--1631},
url = {https://www.usenix.org/conference/usenixsecurity18/presentation/adi},
publisher = {USENIX Association},
month = aug
}

@inproceedings{MSE,
  title     = {Comparing Kullback-Leibler Divergence and Mean Squared Error Loss in Knowledge Distillation},
  author    = {Kim, Taehyeon and Oh, Jaehoon and Kim, Nak Yil and Cho, Sangwook and Yun, Se-Young},
  booktitle = {Proceedings of the Thirtieth International Joint Conference on
               Artificial Intelligence, {IJCAI-21}},
  publisher = {International Joint Conferences on Artificial Intelligence Organization},
  editor    = {Zhi-Hua Zhou},
  pages     = {2628--2635},
  year      = {2021},
  month     = {8},
  note      = {Main Track},
  doi       = {10.24963/ijcai.2021/362},
  url       = {https://doi.org/10.24963/ijcai.2021/362},
}

@inproceedings{PUF,
  title={Silicon physical random functions},
  author={Gassend, Blaise and Clarke, Dwaine and van Dijk, Marten and Devadas, Srinivas},
  booktitle={Proceedings of the 9th ACM conference on Computer and communications security},
  pages={148--160},
  year={2002},
  organization={ACM}
}

@inproceedings{PUF_NOISE,
  title={Modeling and evaluating the reliability of a physical unclonable function during device lifetime},
  author={Maes, Roel and van der Leest, Vladimir},
  booktitle={2011 IEEE International Conference on Computer Design (ICCD)},
  pages={480--485},
  year={2011},
  organization={IEEE},
  doi={10.1109/ICCD.2011.6081443}
}

@inproceedings{li2020puf,
  author    = {Dawei Li and Yangkun Ren and Di Liu and Zhenyu Guan and Qianyun Zhang and Yanzhao Wang},
  title     = {{PUF-Based Intellectual Property Protection for CNN Model}},
  booktitle = {Proceedings of the 2020 IEEE International Symposium on Circuits and Systems (ISCAS)},
  year      = {2020},
  pages     = {1--5},
  doi       = {10.1109/ISCAS45731.2020.9180883},
  publisher = {IEEE},
}

@INPROCEEDINGS{ISCAS,
  author={Jiang, Jingdong and Zheng, Yue and Chang, Chip-Hong},
  booktitle={2025 IEEE International Symposium on Circuits and Systems (ISCAS)}, 
  title={PUF-based Edge DNN Model IP Protection with Self-obfuscation and Publicly Verifiable Ownership}, 
  year={2025},
  volume={},
  number={},
  pages={1-5},
  doi={10.1109/ISCAS56072.2025.11044032}}

@INPROCEEDINGS{sram,
  author={Korenda, Ashwija Reddy and Afghah, Fatemeh and Cambou, Bertrand and Philabaum, Christopher},
  booktitle={2019 16th Annual IEEE International Conference on Sensing, Communication, and Networking (SECON)}, 
  title={A Proof of Concept SRAM-based Physically Unclonable Function (PUF) Key Generation Mechanism for IoT Devices}, 
  year={2019},
  volume={},
  number={},
  pages={1-8},
  doi={10.1109/SAHCN.2019.8824887}}

@ARTICLE{ro,
  author={Huang, Zhengfeng and Bian, Jingchang and Lin, Yankun and Liang, Huaguo and Ni, Tianming},
  journal={IEEE Transactions on Computer-Aided Design of Integrated Circuits and Systems}, 
  title={Design Guidelines and Feedback Structure of Ring Oscillator PUF for Performance Improvement}, 
  year={2024},
  volume={43},
  number={1},
  pages={71-84},
  doi={10.1109/TCAD.2023.3301386}}

@article{arb,
title = {Priority Arbiter PUF: Analysis},
journal = {Discrete Applied Mathematics},
volume = {356},
pages = {71-95},
year = {2024},
issn = {0166-218X},
doi = {https://doi.org/10.1016/j.dam.2024.05.013},
url = {https://www.sciencedirect.com/science/article/pii/S0166218X24002026},
author = {Meenakshi Kansal and Animesh Roy and Dibyendu Roy and Srinivasu Bodapati and Anupam Chattopadhyay},
}

@article{variation,
  title={Impact of temperature fluctuations on SRAM-based physically unclonable functions},
  author={Kim, Moon-Seok and Yoo, Sang-Kyung and Kang, Junki and Kim, Sungho},
  journal={International Journal of Information Security},
  volume={24},
  number={3},
  pages={113},
  year={2025},
  publisher={Springer}
}

@INPROCEEDINGS{variation1,
  author={Alheyasat, A. and Torrens, G. and Bota, S. and Alorda, B.},
  booktitle={2020 IEEE International Symposium on Circuits and Systems (ISCAS)}, 
  title={Bit-Cell Selection Analysis for Embedded SRAM-Based PUF}, 
  year={2020},
  volume={},
  number={},
  pages={1-4},
  doi={10.1109/ISCAS45731.2020.9180780}}

@article{distillingknowle,
  title={Distilling the Knowledge in a Neural Network},
  author={Geoffrey E. Hinton and Oriol Vinyals and Jeffrey Dean},
  journal={ArXiv},
  year={2015},
  volume={abs/1503.02531},
  url={https://api.semanticscholar.org/CorpusID:7200347}
}

@inproceedings{learningmodels,
  author    = {Tram{\`e}r, Florian and Zhang, Fan and Juels, Ari and Reiter, Michael K. and Ristenpart, Thomas},
  title     = {Stealing Machine Learning Models via Prediction APIs},
  booktitle = {25th USENIX Security Symposium (USENIX Security 16)},
  year      = {2016},
  pages     = {601--618},
  address   = {Austin, TX, USA},
  month     = aug,
  publisher = {USENIX Association},
  url       = {https://www.usenix.org/conference/usenixsecurity16/technical-sessions/presentation/tramer}
}

@INPROCEEDINGS{reverse,
  author={Hua, Weizhe and Zhang, Zhiru and Suh, G. Edward},
  booktitle={2018 55th ACM/ESDA/IEEE Design Automation Conference (DAC)}, 
  title={Reverse Engineering Convolutional Neural Networks Through Side-channel Information Leaks}, 
  year={2018},
  volume={},
  number={},
  pages={1-6},
  doi={10.1109/DAC.2018.8465773}}

@article{privacypreserving,
  title={Chiron: Privacy-preserving Machine Learning as a Service},
  author={Tyler Hunt and Congzheng Song and R. Shokri and Vitaly Shmatikov and Emmett Witchel},
  journal={ArXiv},
  year={2018},
  volume={abs/1803.05961},
  url={https://api.semanticscholar.org/CorpusID:3970945}
}

@inproceedings{zhang2018protecting,
  title={Protecting intellectual property of deep neural networks with watermarking},
  author={Zhang, Jialong and Gu, Zhongshu and Jang, Jiyong and Wu, Hui and Stoecklin, Marc Ph and Huang, Heqing and Molloy, Ian},
  booktitle={Proceedings of the 2018 ACM SIGSAC Conference on Computer and Communications Security},
  pages={159--172},
  year={2018},
  organization={ACM}
}

@ARTICLE{instable,
  author={Holcomb, Daniel E. and Burleson, Wayne P. and Fu, Kevin},
  journal={IEEE Transactions on Computers}, 
  title={Power-Up SRAM State as an Identifying Fingerprint and Source of True Random Numbers}, 
  year={2009},
  volume={58},
  number={9},
  pages={1198-1210},
  doi={10.1109/TC.2008.212}}

@article{noise,
    author = {Bishop, Chris M.},
    title = {Training with Noise is Equivalent to Tikhonov Regularization},
    journal = {Neural Computation},
    volume = {7},
    number = {1},
    pages = {108-116},
    year = {1995},
    month = {01},
    issn = {0899-7667},
    doi = {10.1162/neco.1995.7.1.108},
    url = {https://doi.org/10.1162/neco.1995.7.1.108},
    eprint = {https://direct.mit.edu/neco/article-pdf/7/1/108/812990/neco.1995.7.1.108.pdf},
}










\end{document}